\documentclass[%
 aip,
cp,  
 amsmath,amssymb,
 reprint,%
]{revtex4}

\usepackage{graphicx}
\usepackage{dcolumn}
\usepackage{bm}

\usepackage[utf8]{inputenc}
\usepackage[T1]{fontenc}
\usepackage{mathptmx}

\begin{document}

\title{Thermal and Pressure Ionization in Warm, Dense MgSiO$_3$ Studied with First-Principles Computer Simulations}

\author{Felipe Gonz\'alez-Cataldo}
\email[Corresponding author: ]{f\_gonzalez@berkeley.edu}
\affiliation{Department of Earth and Planetary Science, University of California, Berkeley, California 94720, USA}
\author{Burkhard Militzer}
\affiliation{Department of Earth and Planetary Science, University of California, Berkeley, California 94720, USA}
\affiliation{Department of Astronomy,                   University of California, Berkeley, California 94720, USA}

\date{\today} 

\begin{abstract}
  Using path integral Monte Carlo and density functional molecular
  dynamics (DFT-MD) simulations, we study the properties of MgSiO$_3$
  enstatite in the regime of warm dense matter. We generate a
  consistent equation of state (EOS) that spans across a wide range of
  temperatures and densities (10$^4$--10$^7$ K and 6.42--64.16 g
  cm$^{-3}$). We derive the shock Hugoniot curve, that is in good
  agreement with the experiments. We identify the boundary between the
  regimes of thermal ionization and pressure ionization by locating
  where the internal energy at constant temperature attains a minimum
  as a function of density or pressure. At low density, the internal
  energy decreases with increasing density as the weight of free
  states changes. Thermal ionization dominates. Conversely, at high
  density, in the regime of pressure ionization, the internal energy
  increases with density. We determine the boundary between the two
  regimes and show that the compression maximum along the shock Hugoniot
  curve occurs because K shell electrons become thermally ionized
  rather than pressure ionized.
\end{abstract}  

\maketitle


\section{Introduction}

Enstatite, MgSiO$_3$, is one of the most abundant minerals on Earth's
crust and one of the few silicate minerals that has been observed in
crystalline form outside the Solar System~\cite{Molster2001}.
Therefore, it has been identified as one the primary building blocks
of planetary formation~\cite{Valencia2010,Bolis2016} and it is commonly
used as a representative mantle material in exoplanet
modeling~\cite{Seager2007,Valencia2009,Valencia2010,Wagner2012}. The
equation of state (EOS) of silicate materials at high pressures and
temperatures is fundamental to develop models of planetary
formation~\cite{Gonzalez-Cataldo2016} and to interpret results from
shock compression experiments~\cite{Fratanduono2018}. In planetary
collision events, for example, the amount of heat generated during the
shock can be sufficient to melt and even vaporize these
minerals~\cite{Kraus2012}, and an accurate EOS is needed to describe
these processes quantitatively.
On the other hand, precise first-principles computer simulations of silicates can guide the
design of inertial confinement fusion (ICF) experiments~\cite{Lindl1995,Gaffney2018,ZhangBN2019}
under
conditions where the K and L shell electrons are gradually ionized,
which is challenging to predict accurately with analytical EOS models.

The properties of the liquid phase of MgSiO$_3$ at high pressure have
recently been explored in laser-driven shock compression
experiments~\cite{Spaulding2012,Bolis2016,Fratanduono2018}, and
suggest a metallic-like behavior over a wide range of
pressure-temperature conditions. Similar findings were reported with
first-principles calculations~\cite{Soubiran2018} that demonstrated
that super-Earth planets can generate magnetic fields within their
mantles.

In this work, we combine path integral Monte Carlo (PIMC) and density
functional theory molecular dynamics (DFT-MD) simulations to study the
ionization in warm, dense MgSiO$_3$. We distinguish the regimes of
thermal ionization and pressure ionization, and show that the shock
Hugoniot curve crosses from one regime into the other. The compression
maximum along the Hugoniot curve is found to fall within the regime of
thermal ionization. The K shell electrons of all three nuclei
contribute to the compression maximum.

\section{Methodology}

Rigorous discussions of the PIMC~\cite{Ce95,Ce96,MilitzerDriver2015,MilitzerPollockCeperley2019}
and DFT-MD~\cite{Car1985,Payne1992,Marx2009} methods have been
provided in previous works, and the details of our simulations have
been presented in some of our previous
publications~\cite{Mi99,DriverNitrogen2016,Soubiran2019}.
Here we summarize the methods and provide the simulation parameters
specific to simulations of MgSiO$_3$ plasma. The general idea of our
approach is to perform simulations along isochores at high
temperatures ($T \geqslant 10^6$ K) using PIMC and at low temperatures
($T \leqslant 10^6$ K) using DFT-MD simulations.  We find the two
methods produce consistent results at overlapping temperatures~\cite{Zhang2018,ZhangBN2019}.

For PIMC simulations, we use the CUPID
code~\cite{MilitzerThesis,Mi01,Mi09} with Hartree-Fock
nodes~\cite{ZhangSodium2017, Driver2017, Driver2018}.  For DFT-MD simulations, we
employ Kohn-Sham DFT simulation techniques as implemented in the
Vienna Ab initio Simulation Package (VASP)~\cite{VASP} using the
projector augmented-wave (PAW) method~\cite{PAW,Kresse1999}, and
molecular dynamics is performed in the NVT ensemble, regulated with a
Nos\'e thermostat.  Exchange-correlation effects are described using
the Perdew, Burke, and Ernzerhof~\cite{PBE} (PBE) generalized gradient
approximation (GGA).  The pseudopotentials used in our DFT-MD
calculations freeze the electrons of the 1s orbital, which leaves 10,
12, and 6 valence electrons for Mg, Si, and O atoms, respectively.
Electronic wave functions are expanded in a plane-wave basis with an
energy cut-off as high as 7000 eV in order to converge the total
energy. Size convergence tests with up to a 65-atom simulation cell at
temperatures of 10\,000 K and above, indicate that pressures are
converged to better than 0.6\%, while internal energies are converged
to better than 0.1\%.  We find, at temperatures above 500\,000 K, that
15-atom supercells are sufficient to obtain converged results for both
energy and pressure, since the kinetic energy far outweighs the
interaction energy at such high
temperatures~\cite{Driver2012,Driver2015,Driver2018}. The number of
bands in each calculation was selected such that orbitals with
occupation as low as $10^{-4}$ were included, which requires up to
14\,000 bands in an 15-atom cell at $2\times10^6$ K and two-fold
compression.  All simulations are performed at the $\Gamma$ point of
the Brillouin zone, which is sufficient for high temperature fluids,
converging total energy to better than 0.01\% compared to a grid of
$k$-points.

\section{Results}

\begin{figure}[!hbt]
    \centering
    \includegraphics[width=8cm]{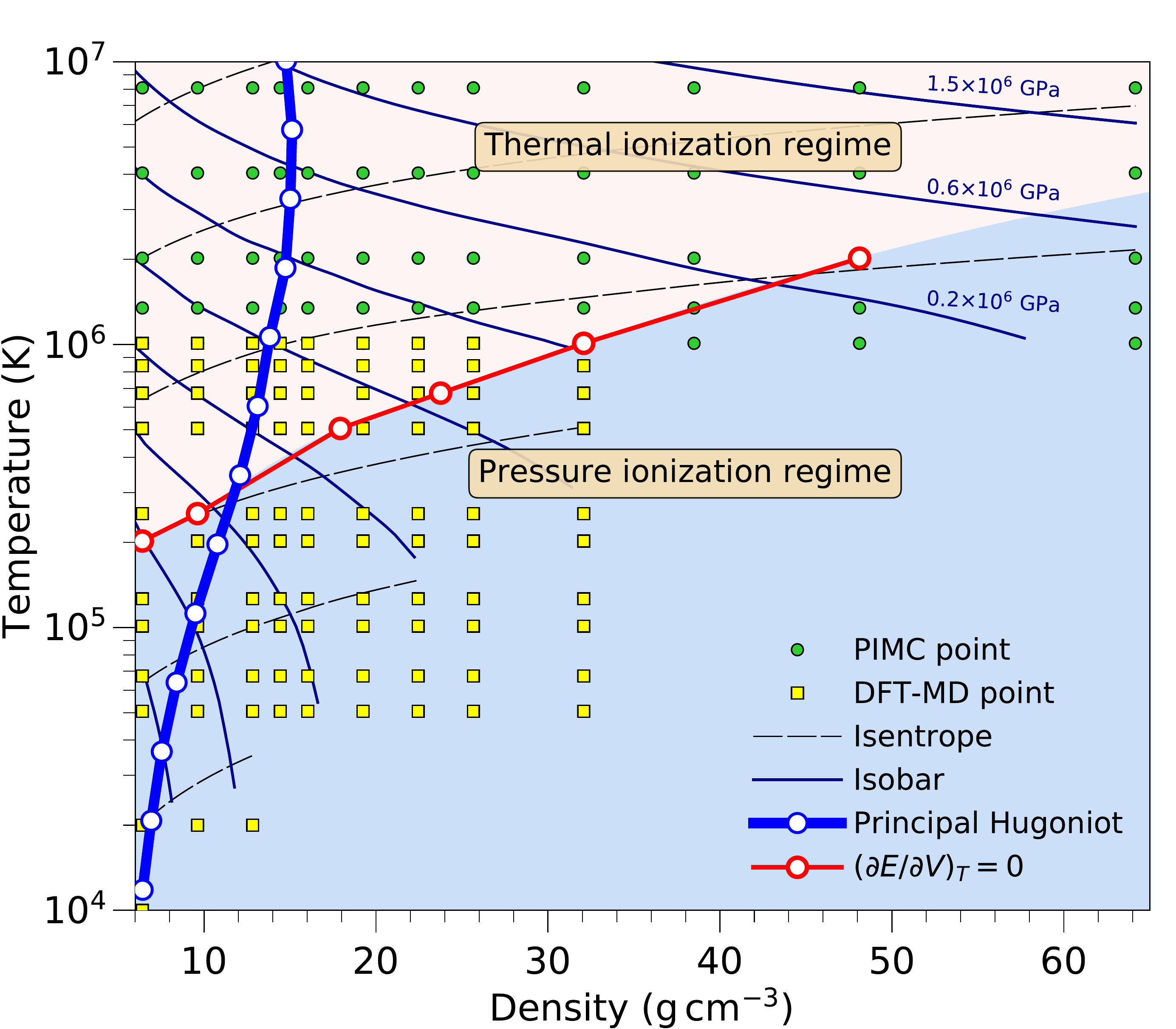}
    \includegraphics[width=8cm]{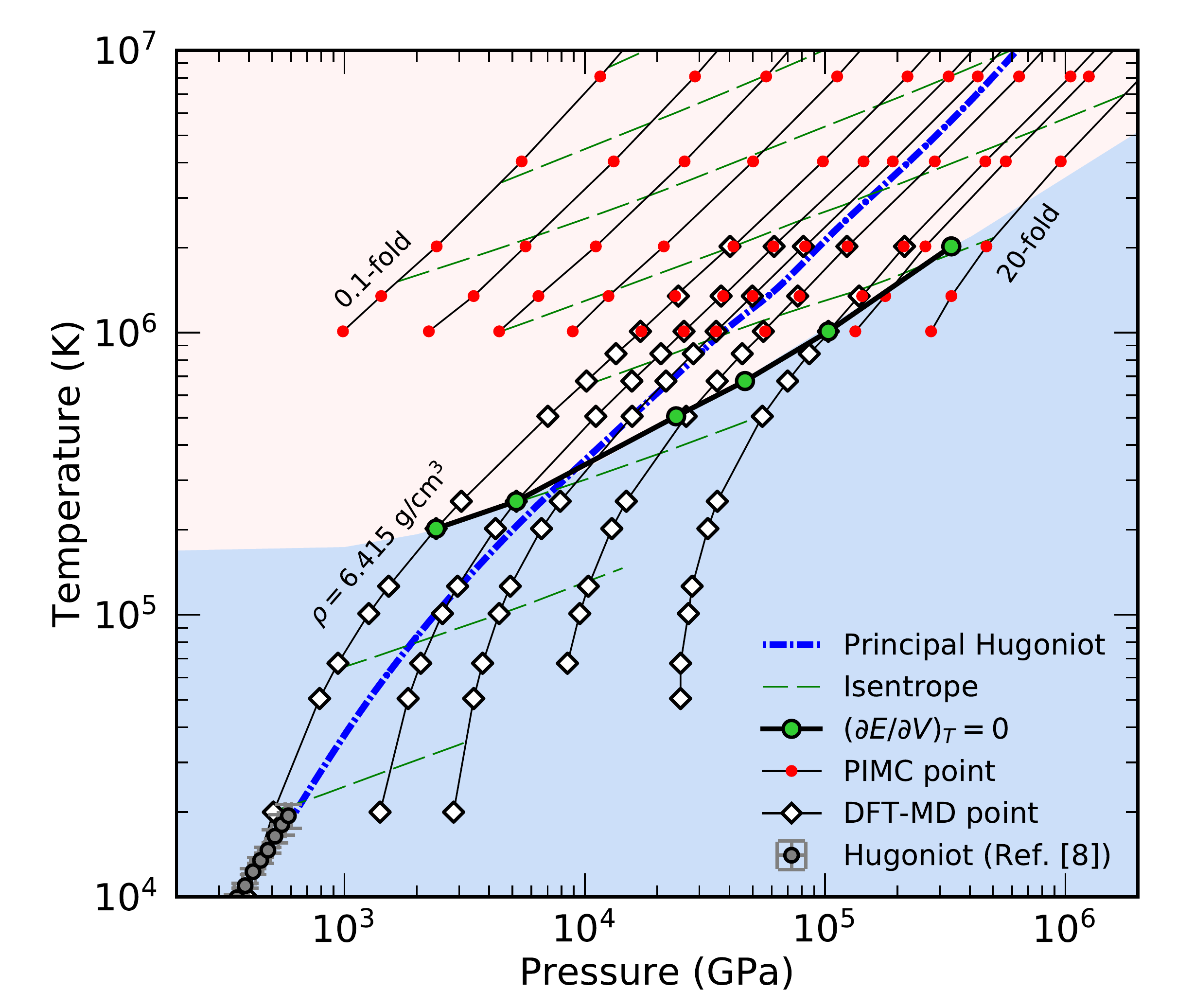}
    \caption{Conditions of our PIMC and DFT-MD and simulations along with computed
isobars and isentropes.  Densities from 0.1-fold (0.321 g cm$^{-3}$) to 20-fold
(64.16 g cm$^{-3}$) and temperatures from $10^4$--$10^7$ K were
considered. The thick blue line
shows the principal Hugoniot curve of MgSiO$_3$ derived from our
simulations~\cite{GonzalezMilitzer2019},
using an initial density of $\rho_0=3.207911$ g cm$^{-3}$
($V_0=51.965073$~\AA/f.u.).
    }\label{fig:Tvsrho}
\end{figure}

In Fig.~\ref{fig:Tvsrho}, we show our EOS data points from
Ref.~\cite{GonzalezMilitzer2019} in temperature-density and temperature-pressure
space.  Computations were performed for a series of densities and
temperatures, ranging from 6.42--64.16 g$\,$cm$^{-3}$ and
$10^4$--$10^{7}$ K, respectively.  To provide a guide for future ramp
compression experiments, we also plot different isentropes, derived
from the relationship
$\left.\frac{dT}{dV}\right|_S=-T\left.\frac{dP}{dT}\right|_V/\left.\frac{dE}{dT}\right|_V$~\cite{Mi09}.
We find that the slope of the isentropes does not strongly depend on
temperature, even though we compare conditions with differing degrees
of ionization. Our results imply that the temperature rise with
pressure along the isentropes approximately follows a power law,
$T\propto P^\alpha$, with an exponent $\alpha=0.309$ below $10^6$ K,
increasing only up to $\alpha=0.399$ for temperatures above $10^7$ K.
This provides a simple rule for obtaining isentropic profiles in
MgSiO$_3$ with wide-range validity, without the need of developing
models for other properties of the material, such as the
Mie--Gr\"uneisen model.
%

The Hugoniot curve of MgSiO$_3$ obtained from shock experiments by
Fratanduono~\emph{et al}.~\cite{Fratanduono2018} is plotted along with
our predicted Hugoniot curve on the temperature-pressure plot of
Fig.~\ref{fig:Tvsrho}.  In the experiment, laser-driven shocks were
used to compress enstatite up to 600 GPa, reaching temperatures as
high as $2\times10^4$ K. This allowed to explore the solid and liquid
regimes of MgSiO$_3$ and obtain a continuous measurement of the
principal Hugoniot curve.  We observe that our Hugoniot curve agrees
very well with the experimental data.

\begin{figure}
    \centering
    \includegraphics[width=8cm]{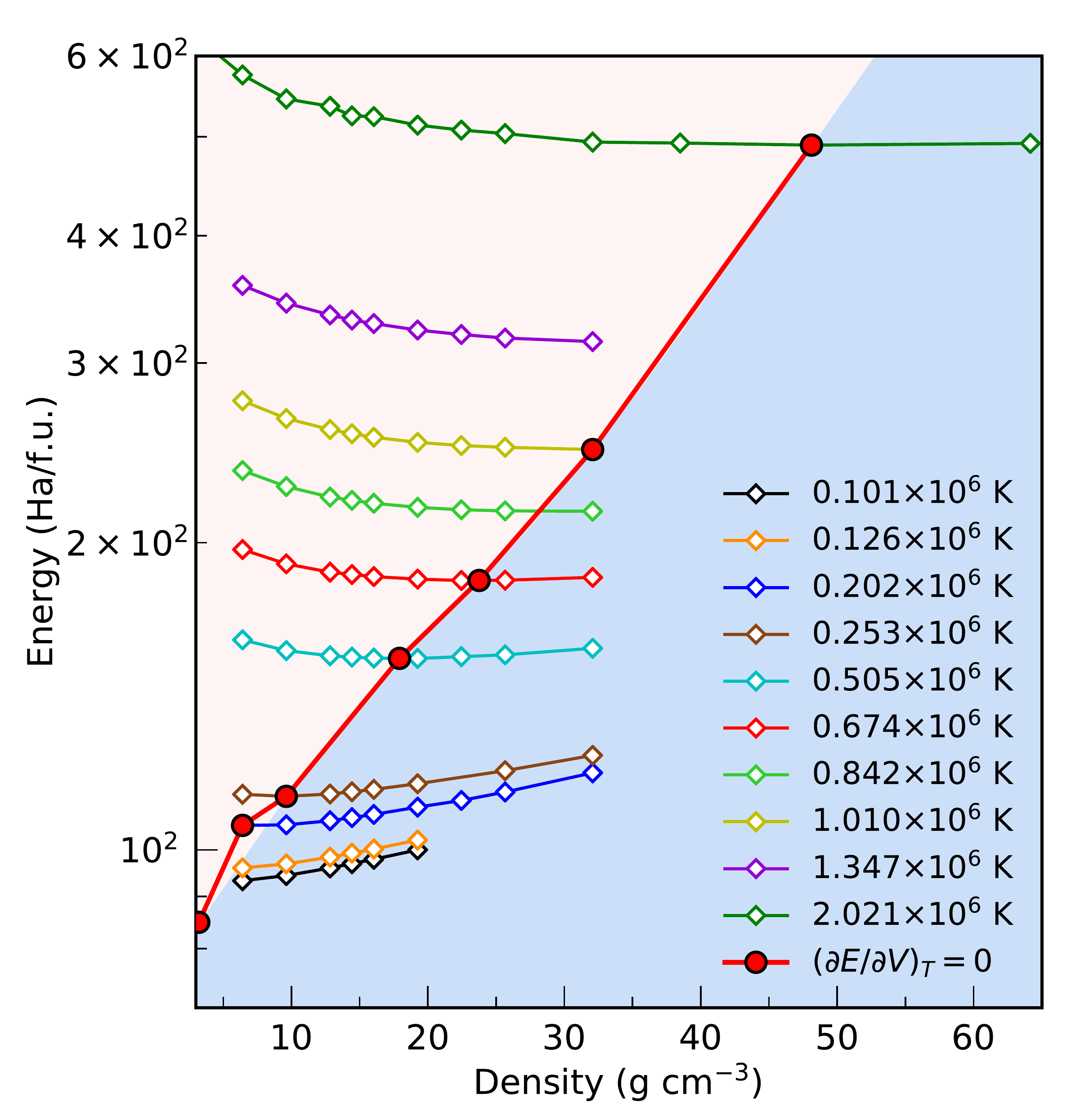}
    \includegraphics[width=8cm]{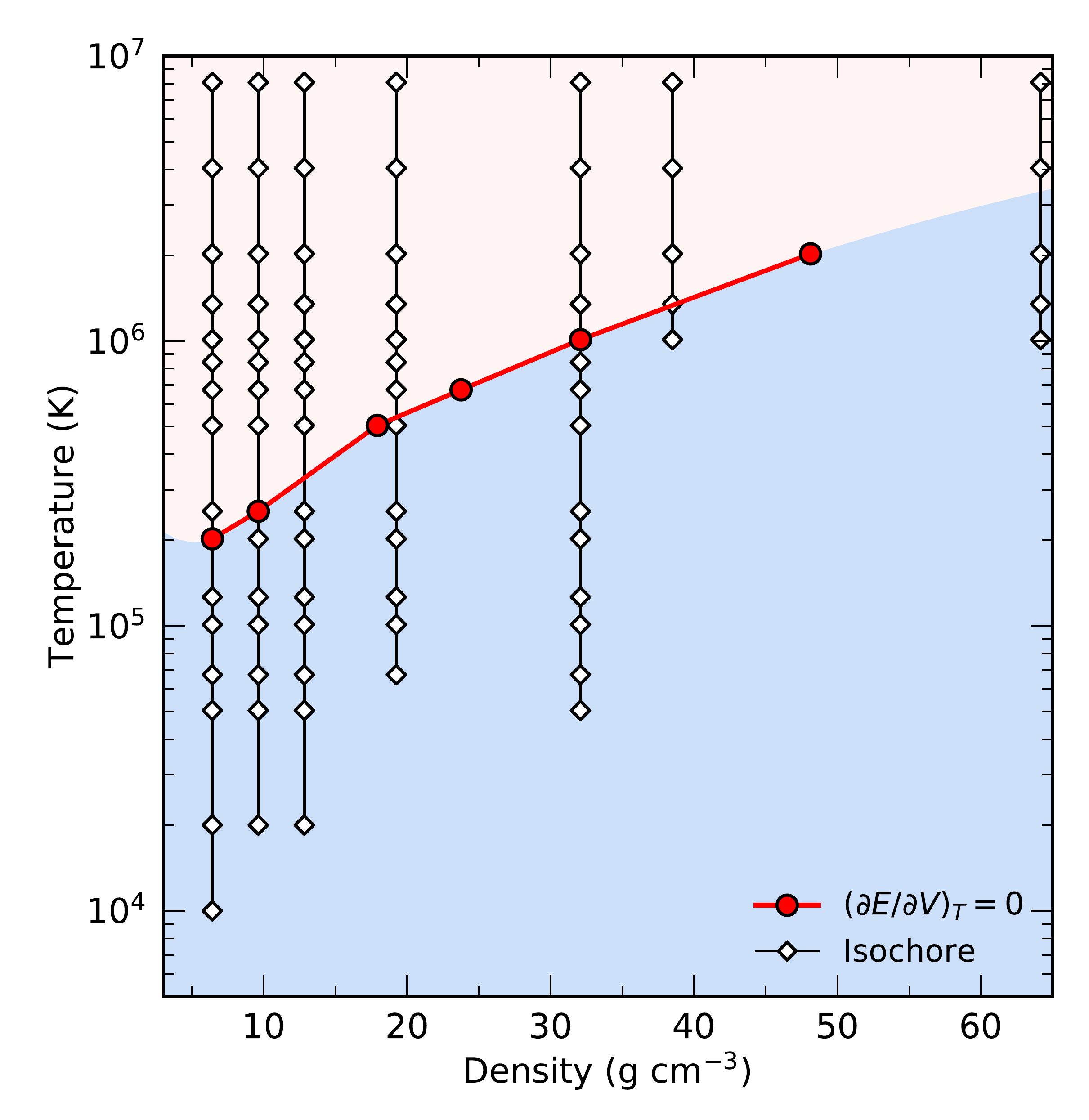}
    \caption{Left: Total energy of the MgSiO$_3$ system as a function
      of density along different isotherms. All energies have been
      shifted upwards by $800$ Ha/f.u. in order to plot only positive
      energies on the logarithmic scale. The energy minimum on each
      isotherm, which satisfies the condition in
      Equation~\eqref{eq:CondEmin}, is shown by the solid red
      circles. Right: MgSiO$_3$ isochores showing the same energy
      minima  in the temperature-density space.
}
\label{fig:Evsrho}
\end{figure}

\begin{figure}
    \centering
    \includegraphics[width=8cm]{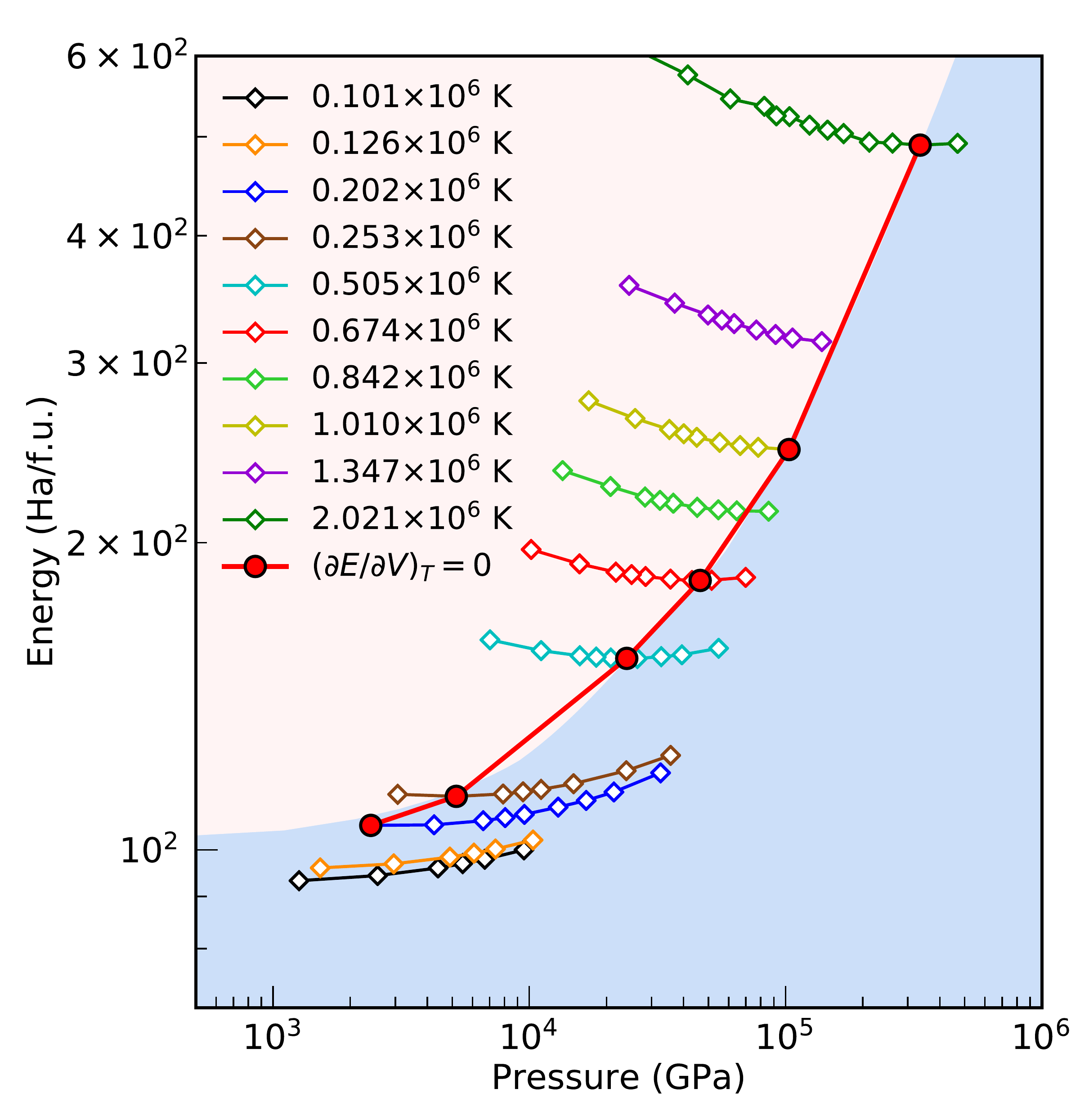}
    \includegraphics[width=8cm]{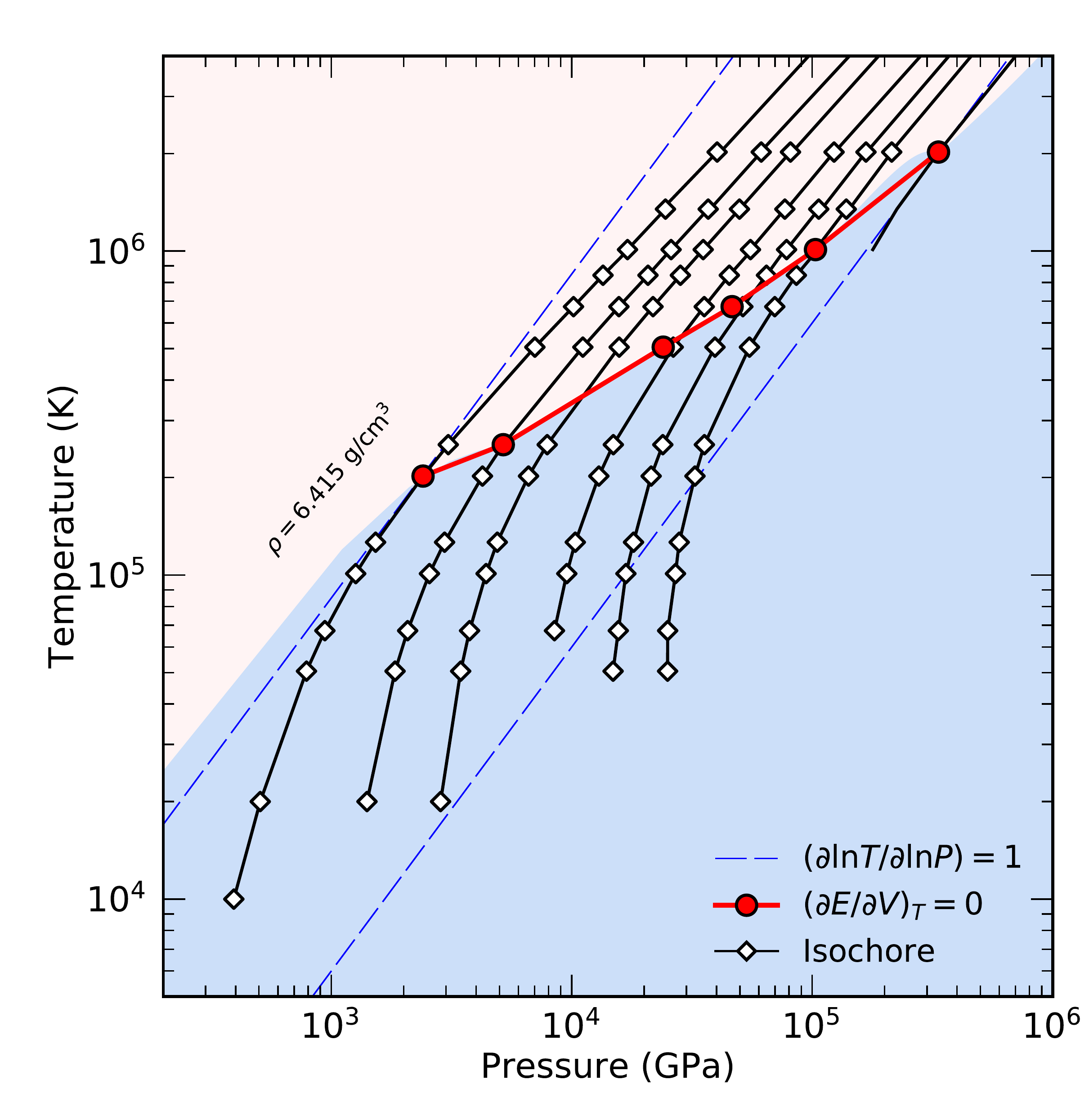}
    \caption{Left: Total energy of the MgSiO$_3$ system as a function
      of pressure along different isotherms. The energies have been
      shifted upwards by $800$ Ha/f.u. in order to plot only positive
      energies in logarithmic scale. The energy minimum,
      which satisfies the condition in 
      Equation~\eqref{eq:CondEmin}, is shown by the solid red circles.
      Right: MgSiO$_3$ isochores showing the same energy minima  in
      the pressure-temperature space, where
      condition~\eqref{eq:CondEmin} can be visualized. Two curves of
      slope equal to one in this $\log T$-$\log P$ plot are shown in
      dashed blue lines.}
\label{fig:EvsPressure}
\end{figure}

In Fig.~\ref{fig:Evsrho}, we plot the internal energy of the system at
fixed temperature as a function of density. Most curves exhibit a
minimum that satisfies the condition,
\begin{equation}\label{eq:ChainRule}
0 = \left.\frac{\partial E}{\partial V}\right|_T \equiv
\left.\frac{\partial E}{\partial P}\right|_T
\left.\frac{\partial P}{\partial V}\right|_T\;.
\end{equation}
Because thermodynamic stability requires $\left.\frac{\partial P}{\partial V}\right|_T < 0$, this conditions is only satsified if  
$\left.\frac{\partial E}{\partial P}\right|_T=0$. From the first law of thermodynamics, $dE=T\,dS-P\,dV$,
it follows that
\begin{equation}
\left.\frac{\partial E}{\partial V}\right|_T =
T\left.\frac{\partial S}{\partial V}\right|_T -P= T\left.\frac{\partial P}{\partial T}\right|_V -P,
\end{equation}
where we have used the Maxwell identity
$\left.\frac{\partial S}{\partial V}\right|_T= \left.\frac{\partial P}{\partial T}\right|_V$. 
Therefore, the condition in Equation~\eqref{eq:ChainRule} is satisfied if
\begin{equation}
0 = \left.\frac{\partial E}{\partial V}\right|_T = T\left[\left.\frac{\partial P}{\partial T}\right|_V -\frac PT\right]= T\left[\beta_V-\frac PT\right] =0,
\end{equation}
where $\beta_V\equiv\left.\frac{\partial P}{\partial T}\right|_V$
is the thermal pressure coefficient. Thus, the energy
attains a mininum along an isotherm if the thermal pressure coefficient satisfies
\begin{equation}\label{eq:CondEmin}
\beta_V = \frac PT \;\; , \;\;{\rm ~or~equivalently~} \left.\frac{\partial \ln P}{\partial \ln T}\right|_V=1 \;\;.
\end{equation}
We visualize this condition in the right panel of
Fig.~\ref{fig:EvsPressure}, where the dashed lines have a constant
slope of 1 in the $\log(T)$-$\log(P)$ space.

For example, along the $T=0.202\times10^6$ K isotherm shown in
Fig.~\ref{fig:Evsrho}, we find an energy minimum around
$\rho\approx 6.42$ g cm$^{-3}$ while the slope
$\left.\frac{\partial \ln P}{\partial \ln T}\right|_V$ becomes 1 at
precisely the same temperature on the corresponding isochore, as we
show  in Fig.~\ref{fig:EvsPressure}. This slope descreases with
increasing temperature, but at some point along the isochore this trend
reverses and the slope starts to increase, converging asymptotically
to 1 in the ideal gas limit at very high temperature.
This condition is trivially fulfilled for an ideal gas, that satisfies
$\left.\frac{\partial E}{\partial V}\right|_T=0$ at any condition.  At
very high temperature, where all atomic species are completely
ionized, we find that MgSiO$_3$ starts to behave like an ideal gas and
the isochores approach a slope of 1.  We confirm that this condition
is satisfied in Fig.~\ref{fig:Tvsrho}, that illustrates how all the
isochores attain a slope of 1 at sufficiently high temperature.

From Equation~\eqref{eq:ChainRule} we infer that, if the energy
attains a minimum as function of density it will also exhibit a
corresponding minimum as a function of pressure, which we illustrate
in the left panel of Fig.~\ref{fig:EvsPressure}.

\begin{figure}
    \centering
    \includegraphics[width=10cm]{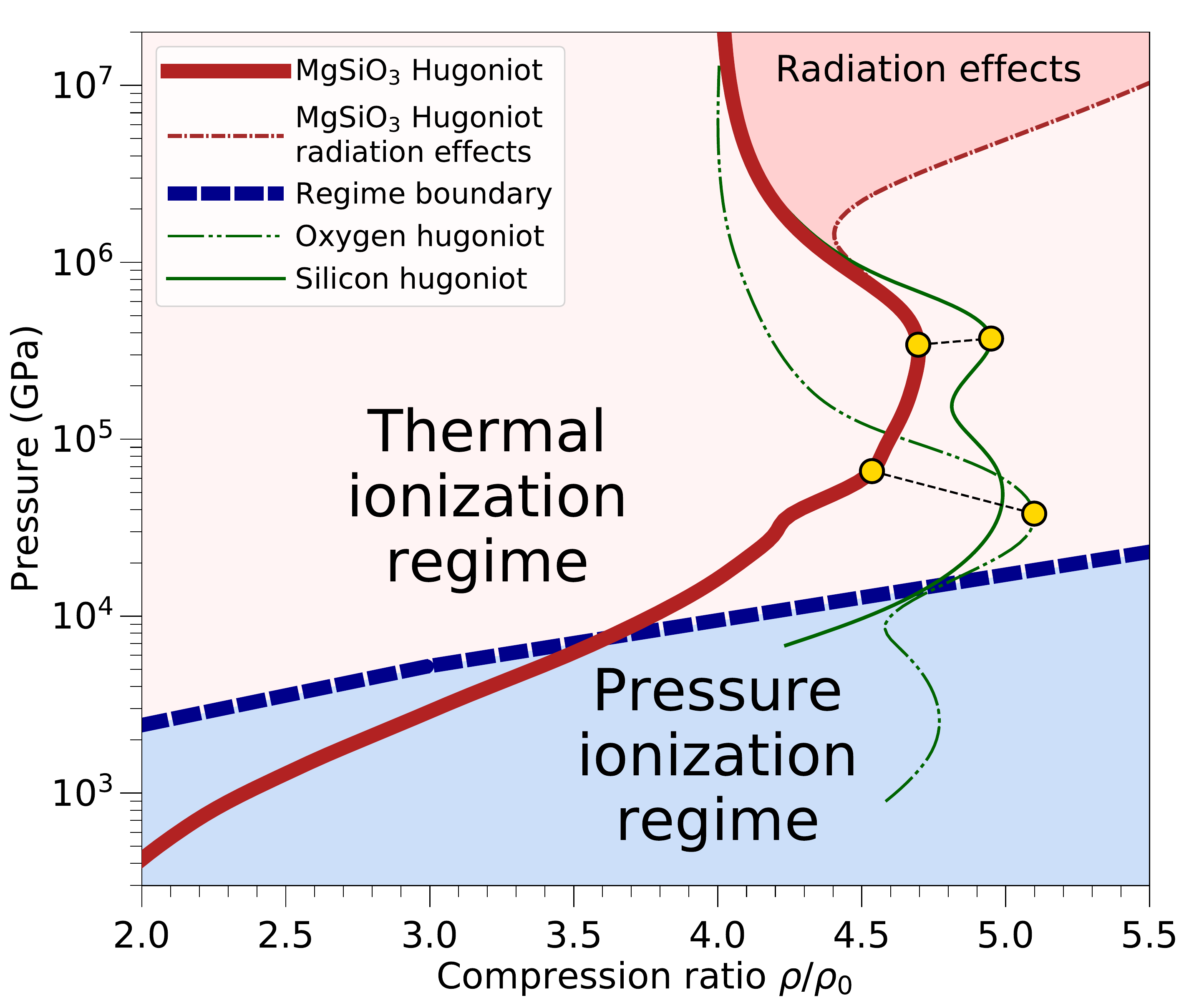}
    \caption{Comparison of the shock Hugoniot curve of MgSiO$_3$,
      oxygen~\cite{Driver2015b}, and
      silicon~\cite{MilitzerDriver2015}. The thermal ionization of K
      shell electrons of Mg, Si, and O species leads to a broad
      maximum in compression. The lower and upper regions of this
      maximum can be associated with the corresponding compression maxima
      of pure oxygen and silicon.}
\label{fig:hug}
\end{figure}

The degree of ionization is controlled by the statistical weights of
free and bound states.  In the limit of low density, where particles
interact weakly, the internal energy increases with decreasing density
because more and more free-particle states become accessible since
their energies scale like $\rho^{2/3}$. Let us assume the energies of bound states do not
change, which is reasonable in the weak-coupling limit. Under these assumptions, the degree of ionization and
the internal energy will always increase if density is decreased at
constant temperature. This regime
is typically described well by the Saha equation~\cite{Ebeling1976}
and we refer to conditions with
$\left.\frac{\partial E}{\partial V}\right|_T>0$ as the {\em thermal
  ionization regime}. Conversely, in the limit of high density, the
internal energy increases with density and we refer to such conditions
as the {\em pressure ionization regime}. Such an internal energy
increase can only occur if the energy of the bound states
increases. There may be several contributing factors including
interaction and Pauli exclusion effects that modify the energy of
bound but also of free states. In some cases, bound states merge with
the continuum of free states. All these effects are expected to
contribute to a rise in internal energy with increasing density, as we
observed in our first-principles simulations at high density. We have
shaded the thermal and pressure ionization regimes in the figures
throughout this article. The condition in
Equation~\eqref{eq:ChainRule} marks the boundary between the two
regimes. From Fig.~\ref{fig:hug}, we learn that the ionization of K
shell electrons, that leads to a compression maximum along the shock
Hugoniot curve, is a thermally driven process.

In the limit of low temperature, we expect the
$\left.\frac{\partial E}{\partial V}\right|_T=0$ line to converge to
the density of ambient MgSiO$_3$ of $\rho_0=3.207911$ g cm$^{-3}$,
which marks a minimum in internal energy. As any standard equation of
state illustrates~\cite{BirchMurnaghan,Vinet1987}, the internal energy
of the solid will rise as the material is compressed or expanded
without changing ionization state of the material.  This is an
example for an energy increase that is not associated with an
noticible increase in the ionization fraction. In general, a density
change will affect the energies of bound and free states if the system
is strongly-coupled. It is only if the bound state energies are
affected more severely that an increase in internal energy is
associated with an increase in the degree of ionization. It is also
possible that higher-lying bound states are affected by pressure
ionization while lower-lying, confined states are not yet. Our energy
criterion is not able to make such distinctions. Nevertheless, it
detects increases in the energy of bound states, which is a necessary
condition for pressure ionization to occur. So, in our view, this
renders our energy criterion a useful approach to distinguish the
regimes of thermal and pressure ionization.

\section{Conclusion}

We have constructed a consistent EOS of MgSiO$_3$ over a wide range of
temperature and density conditions by combining results from DFT-MD and PIMC
simulations in order to bridge the warm dense matter and plasma
regimes~\cite{ZhangCH2018,ZhangBN2019}.
One goal of performing high-precision computer simulations based on first
principles is to guide the design of inertial confinement fusion (ICF)
experiments under conditions where the K and L shell electrons are gradually
ionized, which is challenging to predict accurately with analytical EOS models. 

Here we have introduced a simple thermodynamic criterion based on volume dependence of the internal
energy that allows us to distinguish between the regimes of thermal ionization, $\left.\frac{\partial E}{\partial V}\right|_T>0$, 
and pressure ionization $\left.\frac{\partial E}{\partial V}\right|_T<0$. We conclude that the ionization of the K shell
electrons is a thermally activated rather than pressure driven process under
the condition where the compression maximum occurs on the principal shock
Hugoniot curve.

\begin{acknowledgments}
  This work was in part supported by the National Science
  Foundation-Department of Energy (DOE) partnership for plasma science
  and engineering (grant DE-SC0016248), by the DOE-National Nuclear
  Security Administration (grant DE-NA0003842), and the University of
  California Laboratory Fees Research Program (grant
  LFR-17-449059).
  We thank the authors of Ref.~\cite{Fratanduono2018} for providing the
  full data set from their experiments.
  F.G.-C. acknowledges support from the CONICYT
  Postdoctoral fellowship (grant 74160058). Computational support was
  provided by the Blue Waters sustained-petascale computing project
  (NSF ACI 1640776) and the National Energy Research Scientific
  Computing Center.
\end{acknowledgments}

\bibliography{Gonzalez_Militzer_AIP_v12_arXiv}
\bibliographystyle{aipnum4-2}

\end{document}